\documentclass[aps,prb,preprint,showpacs,floatfix,superscriptaddress]{revtex4}

\usepackage{amsmath}
\usepackage[dvips]{graphicx}
\usepackage{dcolumn}
\usepackage{bm}

\newcommand{\im}{{\rm Im}\,}

\begin{document}

\title{\textit{Ab initio} calculations of mean free paths and stopping powers}
\date{\today}

\author{A. P. Sorini}
\affiliation{Department of Physics, University of Washington, Seattle, WA 98195}

\author{J. J. Kas}
\affiliation{Department of Physics, University of Washington, Seattle, WA 98195}

\author{J. J. Rehr}
\affiliation{Department of Physics, University of Washington, Seattle, WA 98195}

\author{M. P. Prange}
\affiliation{Department of Physics, University of Washington, Seattle, WA 98195}

\author{Z. H. Levine}
\affiliation{National Institute of Standards and Technology, Gaithersburg, Maryland 20899}

\begin{abstract}

 A method is presented for first-principles calculations of inelastic
mean free paths and stopping powers in condensed matter over a broad energy
range.  The method is based on {\it ab initio} calculations of the
dielectric function in the long wavelength limit using a real-space
Green's function formalism, together with extensions to finite momentum
transfer. From these results we obtain the loss function and related
quantities such as optical-oscillator strengths and mean excitation
energies.  From a many-pole representation of the dielectric function
we then obtain the electron self-energy and inelastic mean free paths (IMFP).
Finally using our calculated dielectric function and the optical-data model
of Fern\'andez-Varea {\it et al}.,
we obtain collision stopping powers (CSP) and penetration ranges.
The results are consistent with semi-empirical approaches and
with experiment.

\end{abstract}


\maketitle
\newcommand{\imag}[1]{
{\textrm{Im}{\left[{#1}\right]}}
}

\section{I. Introduction }    

  The effect of inelastic losses on fast electrons has long been 
of theoretical and experimental interest \cite{bethe30,fermi40,fano63},
and continues to be an area of active development
\cite{salvat05,powell99,bichsel93}. 
Theoretical calculations of such losses depend on
the dielectric response of a material over a broad spectrum.
Moreover, calculations of losses at low energies are particularly sensitive
to the excitation spectrum of a material.  While
first-principles approaches have been developed for calculations of
losses at low energies, i.e., up to a few tens of eV \cite{rubio99,soininen03},
these methods are computationally intensive and may be difficult
to implement.  Thus detailed calculations of inelastic
losses have generally been limited to semi-empirical
approaches \cite{dpenn87,salvat05,sternheimer82}, based on
experimental optical data.
On the other hand, experimental data over a sufficiently broad spectrum
are not readily available for many materials of interest.

  In an effort to overcome these difficulties, we present here
a first-principles,
real-space approach for general calculations of inelastic losses.
The approach is applicable to both periodic and aperiodic condensed
matter systems throughout the periodic table.
Our calculations are based on {\it ab initio} calculations of
the complex dielectric function
$\epsilon(\omega)=\epsilon_1(\omega)+i\epsilon_2(\omega)$
in the long-wavelength limit,
together with extensions to finite momentum transfer
\cite{rehr06}.  The calculations of $\epsilon(\omega)$ are carried out
using an all electron, real-space Green's function (RSGF) formalism
as implemented in a generalization of the FEFF8
code \cite{ankudinov98,prangeetal05} for full-spectrum calculations
of optical constants.

We focus in this paper on several physical quantities which characterize
the inelastic interactions
of a fast probe electron, a photo-electron, or other charged particle
 in condensed matter.
These include the {inelastic mean-free-path} (IMFP) and the
{collision stopping-power} (CSP).  Each of these quantities depends on
the complex dielectric function $\epsilon(\omega)$ through the
loss function for a given material $-[{\rm Im}\, \epsilon^{-1}(\omega)] =
\epsilon_2(\omega)/[\epsilon_1(\omega)^2+\epsilon_2(\omega)^2]$,
which is calculated here up to x-ray energies.
The loss function is directly related to the optical oscillator
strength (OOS). From the OOS we obtain values of the mean excitation
energy $I$ which characterizes the distribution of
excitations (e.g., plasmons, particle-hole excitations, etc.).
Recently a comprehensive relativistic treatment of inelastic losses
and scattering within the first Born approximation
has been developed by Fern\'andez-Varea
{\it et al.}~\cite{salvat05}.  Their semi-empirical approach 
requires experimental optical data as input, and is referred to
here as the optical data model (ODM). This approach has
the advantage that calculations
of quantities such as the CSP are reduced to a single quadrature. 
 To facilitate precise comparisons, we have used their formulation
for our CSP calculations, except for the substitution
of our {\it ab initio} dielectric function.
Our approach is therefore referred to as
the ``{\it ab initio} data model" (ADM). 
We have also compared IMFPs calculated using both the ADM
and a one-particle {\it self-energy} approach.


Formally the IMFP and the CSP are related to energy moments of the differential
cross-section (DCS) for inelastic collisions $d\sigma$/$d\omega$
of a fast probe electron (or other charged particle) of initial kinetic
energy $E$ with energy loss $\omega$.
The inverse IMFP is proportional to the zeroth moment of the DCS
\begin{equation}\label{IMFP1}
\frac{1}{\lambda(E)}=n_a\int d\omega\,
\frac{d\sigma(\omega;E)}{d\omega}=n_a\sigma^{(0)}(E)\; ,
\end{equation}
where $n_a=N/V$ is the atomic number density.
Here and elsewhere in this paper we use Hartree atomic units
($m=\hbar=e^2=1$). Thus distances are in 
Bohrs ($a_0\approx0.529$~\AA)\, and energies in Hartrees 
(H $\approx$ 27.2~eV), unless otherwise specified.
The CSP, here denoted by $S(E)$, is proportional to the first moment of the DCS
\begin{equation}\label{SP1}
S(E)=n_a\int \omega\, d\omega \frac{d\sigma(\omega;E)}{d\omega}
    =n_a\sigma^{(1)}(E)\;.
\end{equation}
Since $S(E)=-dE/dx$, this quantity has units of force.
 From an integral of $1/S(E)$ over energy we then obtain the net
penetration range or path length
$R(E)$. Implicit in Eqs.~(\ref{IMFP1}) and (\ref{SP1}) 
are the kinematics of the colliding particles. In this paper we choose
kinematics relevant for electrons, but we could alternatively 
use similar equations to describe protons or other ions by suitably modifing
the domain of integration that defines $d\sigma/d\omega$ (see below).
Regardless of the probe, the sample is characterized by the dielectric
function $\epsilon({\bf q},\omega)$. In this paper we consider 
cubic materials which we approximate as isotropic, i.e.,
in which the dielectric function depends only on the magnitude
of the momentum transfer $q=\left|{\bf q}\right|$. 

The DCS may be considered as the sum of longitudinal (instantaneous Coulomb)
and transverse (virtual photon) contributions, denoted below with subscripts
$L$ and $T$ respectively.
The detailed relativistic form of the relationship 
between each contribution to the DCS and the loss function is obtained
by integrating the  double differential cross-section
(DDCS) over the kinematically allowed values of momentum-transfer 
\cite{salvat05},
\begin{equation}
\frac{d\sigma(\omega;E)}{d\omega}
=\int dq\frac{d\sigma(q,\omega;E)}{dqd\omega}\;,
\end{equation}
where 
\begin{equation}
\frac{d\sigma({q},\omega)}{dqd\omega}=
\frac{d\sigma_L({q},\omega)}{dqd\omega}
+
\frac{d\sigma_T({q},\omega)}{dqd\omega}.
\end{equation}
As an example of how the dielectric function determines the DDCS,
we recall the familiar non-relativistic result:
\begin{equation}\label{eq:ddcs}
\frac{d\sigma(q,\omega)}{dqd\omega}=
\frac{d\sigma_L({q},\omega)}{dqd\omega}=-\frac{1}{2\pi n_a q v^2}\,
\im \epsilon^{-1}(q,\omega).
\end{equation}
%
%
The relativistic analog of Eq.~(\ref{eq:ddcs}) is similar and is given
explicitly in Eqs.~(8) and (9) of Ref.~\cite{salvat05}.

One of the main goals of this work is to calculate mean excitation energies
$I$ and IMFPs for general condensed matter systems over an energy range
up to about
100 keV.  Another goal is to calculate CSPs and penetration ranges
over a range of order 10 MeV.
We compare our results both with other semi-empirical
approaches and with experimental data and tabulations.

\section{II. Model Dielectric Function}

\begin{figure}
\includegraphics[scale=0.35,angle=270]{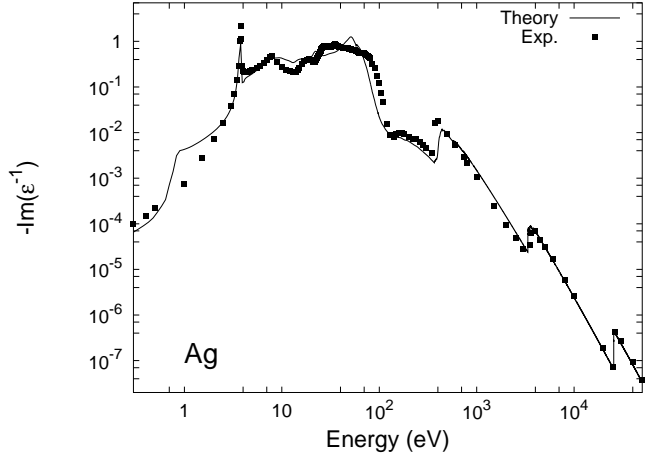}
\caption{\label{fig:eps} Loss function $-[{\rm Im}\, \epsilon^{-1}(\omega)]$
of fcc silver as calculated in this work (solid line) and from
experiment \cite{hagemann75,hagemann74} (dots).}
\end{figure}

Both the IMFP and the CSP can be computed as convolutions of the
momentum-transfer and energy-loss dependent
inverse dielectric function $\epsilon^{-1}({q},\omega)$,
with relativistic weighting functions. The precise details 
of the weighting functions are discussed further below. In this Section we 
discuss the extension of our {\it ab initio} calculation
of $\epsilon({q},\omega)$ in the long wavelength ($q\to0$) limit
to finite $q$ \cite{salvat05,lilj83,dpenn87,ashley79,ashley88}.
In this work $\epsilon(\omega)\equiv\epsilon({0},\omega)$ is calculated
from the UV to x-ray energies using the {\it ab initio} real-space
Green's function code FEFF8OP \cite{prangeetal05,prangetables05},
which sums the contributions to the spectra
over all occupied core and semi-core initial states.  As an example,
the calculated loss function 
for Ag is shown in Fig.~\ref{fig:eps}.

\begin{figure}
\includegraphics[scale=0.35,angle=270]{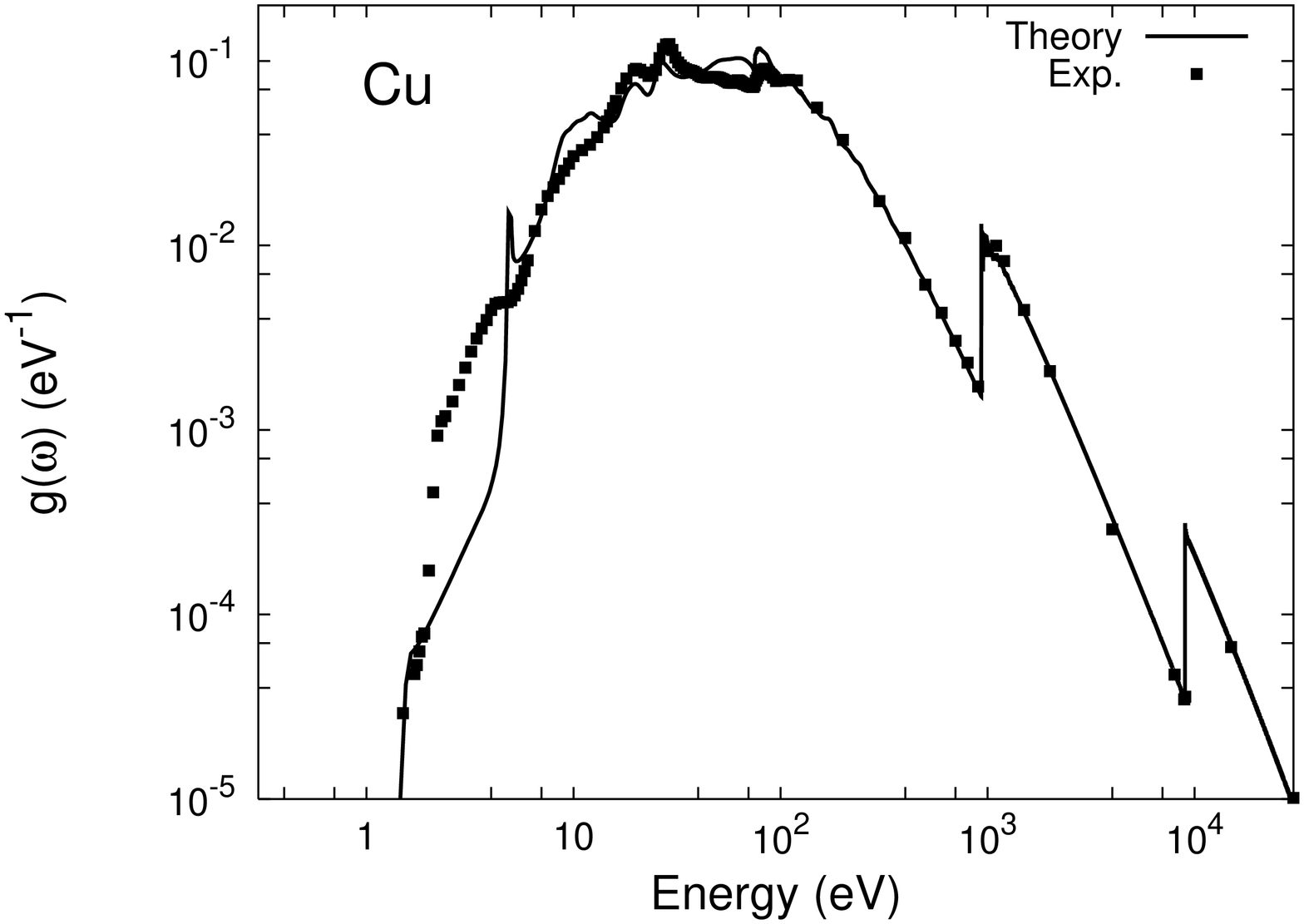}
\includegraphics[scale=0.35,angle=270]{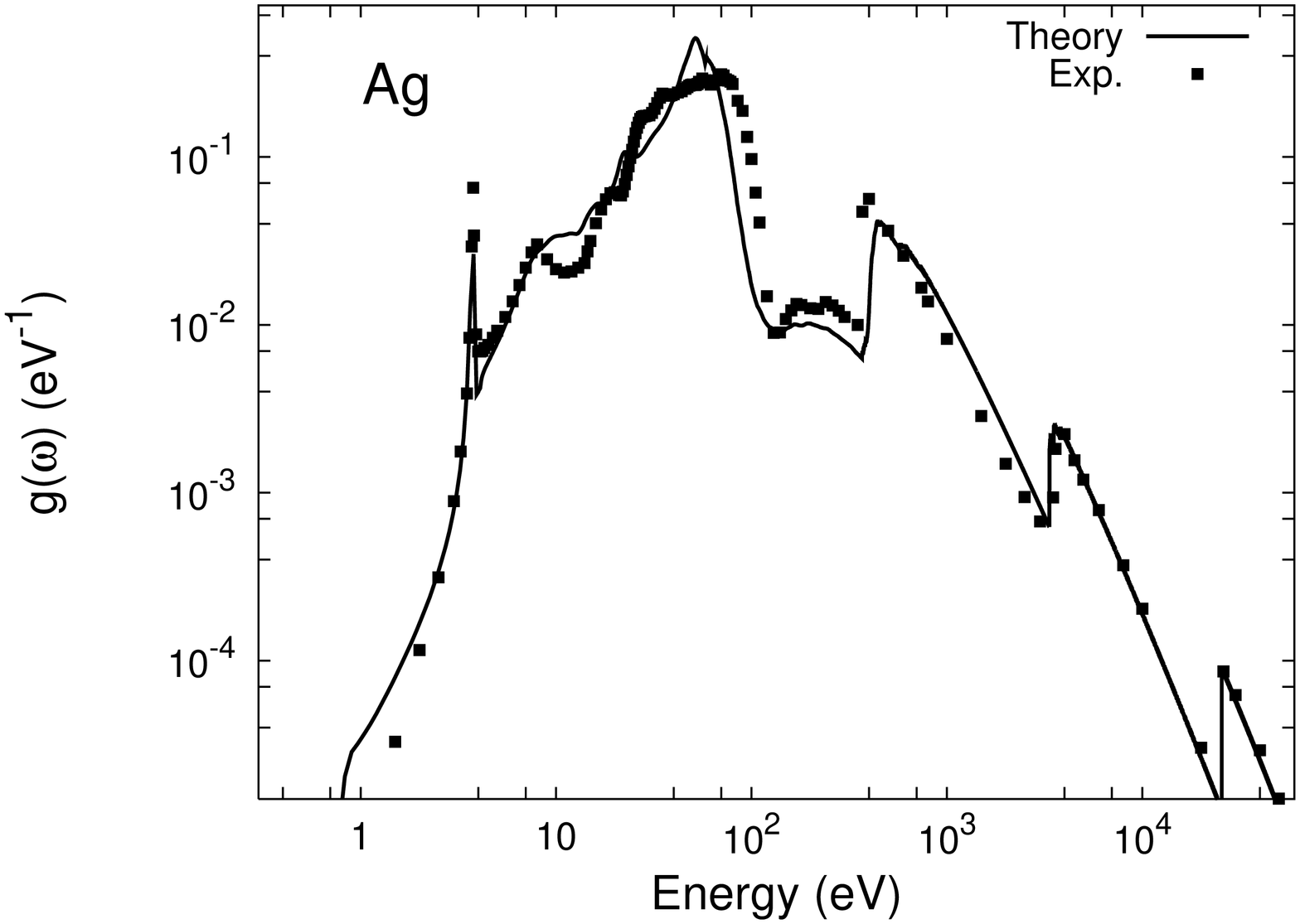}
\includegraphics[scale=0.35,angle=270]{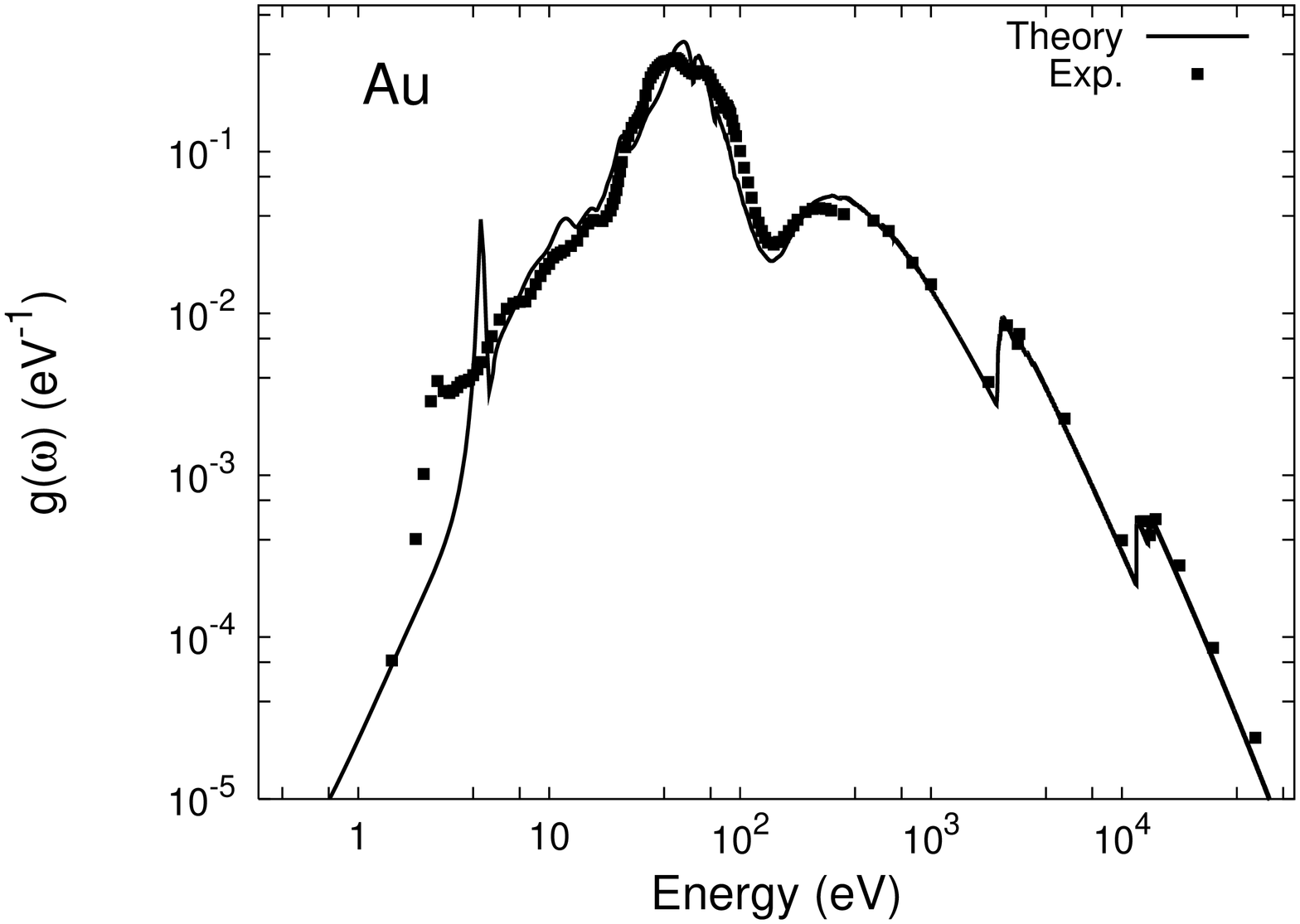}
\caption{\label{fig:oos_cu}Optical oscillator strengths for a) Cu (upper),
b) Ag (middle) and c) Au (lower) as calculated in this work (solid) using the
FEFF8OP code and compared to
experiment \cite{hagemann75,hagemann74} (points).}
\end{figure}

We have chosen to discuss the extension to finite-$q$ in terms of 
the loss function, but we could just as well have 
used the OOS $g(\omega)$,
with differs by a factor proportional to $\omega$, i.e.,
\begin{equation}\label{eqn:oos}
g(\omega)  =
-\frac{2}{\pi}\frac{Z}{\Omega_p^2}\,\omega\, \imag{\epsilon^{-1}(\omega)},
\end{equation}
where $\Omega_p^2=4\pi n_aZ$ is the all-electron plasma frequency.
As an illustration of the quantitative agreement of our approach, three
{\it ab initio} OOS calculations,  spanning a range of atomic numbers,
are compared to experiment in Fig.~\ref{fig:oos_cu}. 
Clearly the approximations in FEFF8 such as the use of
atomic core initial states in the OOS calculation and muffin-tin 
scattering potentials are adequate to yield
good agreement with experiment for UV energies and above.
Additional examples are tabulated on the WWW \cite{sorinitables06}.
For optical frequencies and below, however, the agreement is
only semi-quantitative, but the errors tend to be suppressed in the
OOS due to the overall factor of $\omega$ in Eq.~(\ref{eqn:oos}).
%

A global measure of the excitation spectra is given by
the ``mean excitation energy" $\ln I = \langle\ln \omega\rangle$, where
the ``mean" $\langle\ldots\rangle$ refers to an average with respect to
the OOS weighting function, i.e.,
\begin{equation}
\ln I 
      = \frac{\int d\omega\, g(\omega) \ln \omega }
             {\int d\omega\, g(\omega)}.
\end{equation}
\begin{table}
\caption{Mean excitation energies for several elements as calculated
in this work, and for comparison, results calculated from experimental
\cite{bichsel68,hagemann74}
optical constants, and recommended (ICRU \cite{icru84}) values.}
\label{tab:lni}
\begin{ruledtabular}
\begin{tabular}{cccc}
Element & I {\it theory} (eV) & I {\it expt} (eV) & ICRU (eV) \\ 
Aluminum & 165 & 167 \cite{bichsel68}, 162 \cite{hagemann74}  & 166 \\
Silicon & 174 & 173 \cite{bichsel68} & 173 \\
Copper & 312 & 319 \cite{hagemann74} & 322 \\
Silver & 420 & 382 \cite{hagemann74} & 470 \\
Gold & 662  & 752 \cite{hagemann74} & 790 \\
\end{tabular}
\end{ruledtabular}
\end{table}
The quantity $I$ appears in expressions for the collision stopping power,
as shown in Sec.\ V.\ below. In Table \ref{tab:lni} 
theoretical values of ${I}$, as calculated from our OOS spectra, are compared
with those calculated from experimental optical constants \cite{hagemann74},
and also with internationally
recommended (ICRU) values for several elements.
For low $Z$ elements, the theoretical values of $I$ are clearly 
in good agreement with measured values.
For high $Z$ elements $I$ is predicted by Thomas-Fermi models
to be proportional to $Z$.
 The proportionality constant can be determined experimentally to give
a semi-empirical ``rule of thumb" $I\approx 10\, Z$ (eV). In the high Z
regime the agreement between theory and experiment appears to be only
semi-quantitative, but it should be mentioned that the ratio of $I$ to $Z$
is not in fact constant but rather it varies from around $9$ to around $11$
and has been measured to be as low as $8.32$ for lanthanum \cite{bichsel92}.
Furthermore, only the logarithm of $I$ is needed for the determination of
physical quantities, and typical errors in $\ln I$ from the values in
in Table I are only a few percent.  For example, for the case of $100$ keV
electrons in gold, the error in stopping powers calculated using experimental
versus theoretical values of $I$ is only around $2$\%. 

\begin{figure}
\includegraphics[scale=0.35,angle=270]{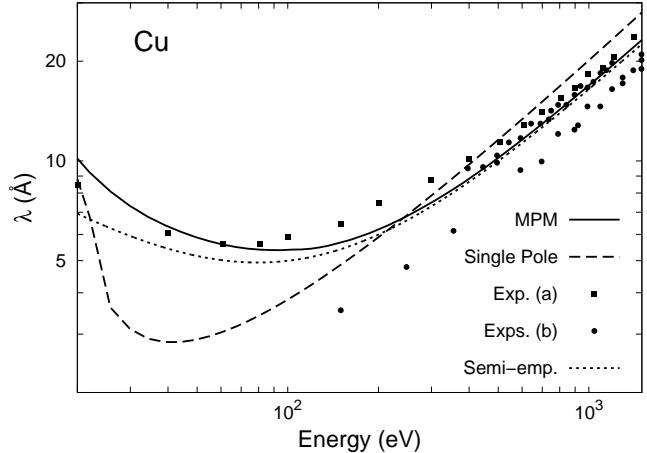}
\caption{\label{fig:imfp_cu}
Inelastic mean free paths for copper calculated using the same {\it ab initio}
dielectric function as the basis of two different theoretical models:
The many-pole self-energy (MPM) model of Eq.~(\ref{IMFP2}) and the single-pole
self-energy model (described in the text). These theoretical results are
compared to: Exp.~(a) \cite{werner} (squares), Exps.~(b) (circles, the references for Exps.~(b) are given in Ref.~\cite{powell99}), and a semi-emperical 
curve which is described in Ref.~\cite{powell99}.}   
%
\end{figure}

In the IMFP calculations presented here we consider two different extensions
to finite $q$, as described below. Since our calculations show that
both lead to similar results for the IMFP, we only present calculations of
the CSP with one of these extensions. However, all our calculations
use the same full-spectrum calculations of $\epsilon(\omega)$. 
%
%
%
%
For our IMFP calculations, we have used the many-pole representation
of the dielectric function of Ref.~\cite{rehr06}, i.e.
we approximate our calculated loss function
 $-{\rm Im}\, [\epsilon^{-1}(\omega)]$ as a sum
of many (typically of order $100$) discrete poles.
%
%
%
This ``many-pole" model, denoted by $\epsilon^{-1}_N(\omega)$,
has the standard analytic form for dielectric response,
\begin{equation}\label{epsJ}
\epsilon^{-1}_N(\omega)=1+\sum_{j=1}^{N} g_j\frac{\omega_j^2}
{\omega^2-\omega_j^2+i\omega\delta},
\end{equation}
where $\delta\omega$ is a small damping term, comparable to
the pole separations.
Fig.~\ref{fig:imfp_cu} compares the IMFP for Cu as calculated using both
our many-pole model and a single plasmon-pole model \cite{lundqvist67}.
This single-pole model is essentially an
Einstein-model for the response in which excitations (for a given
momentum transfer $q$) occur at the plasmon excitation energy
$\omega_q$. Thus the single-pole model is a
special case of the many-pole model in which all but one of the weights
$g_j$ appearing in Eq.~(\ref{epsJ}) are set to zero. 
%
%
In our many-pole representation, the parameters $\omega_j$ are taken to be
evenly spaced along the energy-loss axis, and 
the weights $g_j$ are fixed by matching our 
calculation of $\imag{\epsilon^{-1}(\omega)}$ 
according to the formula
\begin{equation}
g_j=-\frac{2}{\pi}\frac{1}{\omega_j^2}\int_{\Delta_j}d\omega\,\omega\,
 \imag{\epsilon^{-1}(\omega)},
\end{equation}
where the integration region $\Delta_j$ is from $(\omega_j+\omega_{j-1})/2$
to $(\omega_j+\omega_{j+1})/2$ and the similarity with Eq.~(\ref{eqn:oos}) is 
apparent.
Finally, the extension to finite $q$ is obtained by shifting the
pole locations via the substitution \cite{lundqvist67}
\begin{equation}\label{lund}
\omega_j^2\quad\to\quad\omega_j^2+\frac{{v_F}^2q^2}{3}+\frac{q^4}{4}\;,
\end{equation}
where $v_F=k_F/m$ is the Fermi velocity as calculated at the mean
interstitial electron density from the FEFF8 code.
Further details of our approach, though not essential to our discussion
here, are given in Refs.~\cite{rehr06},
\cite{prangetables05} and \cite{kasetal05}.
The above substitution is sufficient to induce the 
so-called ``Bethe ridge" for large momentum-transfer where the loss 
function is peaked about the point $\omega=q^2/2$. 
In other words, our model for large $q$ satisfies the approximate relation
\begin{equation}\label{ridge}
-\imag{ \epsilon^{-1}({q},\omega)}
\approx\pi\Omega_p^2\delta(\omega^2-Q^2),
\end{equation}
where
$Q\equiv q^2/2$. Consequently the above extension can be regarded as an
interpolation formula between small and large $Q$.

For CSP calculations, our aim here is to replace experimental
optical data (which is used as input in the ODM of Ref.~\cite{salvat05})
with theoretical ``optical data" from our {\it ab initio} calculation of
$\epsilon(\omega)$, i.e., with an {\it ab initio} data model (ADM).
Thus for consistency we follow the formulation of Ref.~\cite{salvat05} 
as closely as possible in comparisons with their CSP results,
In particular we have also implemented their delta-oscillator
\cite{lilj83,penelope} extension to finite $q$ for our CSP calculations.
For non-relativistic probe electrons
the delta-oscillator model extends $\epsilon(\omega)$ to 
finite $q$ according to the relation
\begin{eqnarray}\label{ridge2}
 - \imag{\epsilon^{-1}({q},\omega)} &=&
   \pi\Omega_p^2\frac{Z(Q)}{Z}\delta(\omega^2-Q^2) 
\nonumber
\\
&-& \imag{\epsilon^{-1}(\omega)} \theta(\omega-Q) \;,
\end{eqnarray}
where $Z(Q)$ is the number of electrons that contribute to the
zero momentum-transfer sum-rule with upper energy limit $Q$,
\begin{equation}
Z(Q)=-\frac{2Z}{\pi\Omega_p^2}\int_0^Q d\omega\, \omega\,
\imag{\epsilon^{-1}(\omega)}\;.
\end{equation}
Because $Z(Q)$ approaches $Z$ for large $q$ we see that
the extension to finite $q$ in Eq.~(\ref{ridge2})
gives the Bethe ridge in much the same way as
that of Eq.~(\ref{ridge}).
Although the finite-$q$ extension algorithms in this paper
differ somewhat, we do not expect our non-relativistic results to depend 
significantly on the details. The reason is that both algorithms reduce to
the correct long-wavelength limit for low $q$, and both give
the correct Bethe ridge dispersion for high $q$.
This expectation is supported by the IMFP results 
in Sec.\ IV. Moreover our results for the $q$ dependence are roughly
consistent with the explicit real space calculations of
$ - \imag{\epsilon^{-1}({q},\omega)}$ at finite $q$ of Soininen
et al. \cite{soininen05}.

\section{III. Electron Self-Energy}

Inelastic losses in the propagation of a fast charged particle can be
expressed in terms of one-particle {\it self-energy} $\Sigma(E)$.
This complex-valued quantity is a dynamically screened exchange-correlation
contribution to the quasi-particle energy-momentum relation
\begin{equation}
E=\frac{{p}^{2}}{2}+\Sigma(E),
\end{equation}
where $p$ is the quasi-particle momentum.  Our approach for calculating
$\Sigma(E)$ is based on
the ``$GW$" approximation of Hedin \cite{hedin69}, together with
our many-pole representation of the dielectric function,
as summarized above \cite{rehr06}.  In the GW method the
vertex-corrections to the electron self-energy are neglected,
yielding an expression for $\Sigma(E)$ in terms of
the electron propagator $G$ and the screened Coulomb potential acting
on an electron $W$,  i.e.,
%
%
%
\begin{equation}\label{sigma}
\Sigma(E)=i\int\frac{d\omega}{2\pi}e^{-i\omega\eta}G(E-\omega)W(\omega),
\end{equation}
where $\eta$ is a positive infinitesimal and spatial indices 
$({\bf x, x'})$ have been suppressed for clarity.
%
%
Within the RSGF approach, the propagator $G$ is calculated using a
multiple-scattering expansion $G=G_0+G_0 t G_0 +\cdots$ \cite{ankudinov98}.
However, for simplicity in this work, we neglect the multiple-scattering
terms (which would give rise to fluctuations in the self-energy)
and simply use the free propagator $G_0$ for a homogeneous electron gas
at the mean interstitial density.  Then the screened Coulomb
interaction for a spatially homogeneous system can be obtained from the 
Fourier transform 
$W({\bf q},\omega)={\cal F}\left[W({\bf x-x'},t)\right]$
and can be expressed in terms of the Coulomb potential $v_{q}=4\pi$/$q^2$
and the dielectric function
$\epsilon({\bf q},\omega)$ 
 as:
\begin{equation}
\label{scrncoul}
W({\bf  q},\omega)= \epsilon^{-1}({\bf  q},\omega)\, v_{q}.
\end{equation}
The calculations of $\Sigma(E)$ are then  carried out using the
many-pole representation of Eq.~(\ref{epsJ}) and (\ref{lund}).
With this homogeneous model,
our calculated $\Sigma(E)$ is then the average self-energy
in the system.  Further details are given in Ref.~\cite{kasetal05}.



\begin{figure}
\includegraphics[scale=0.35,angle=270]{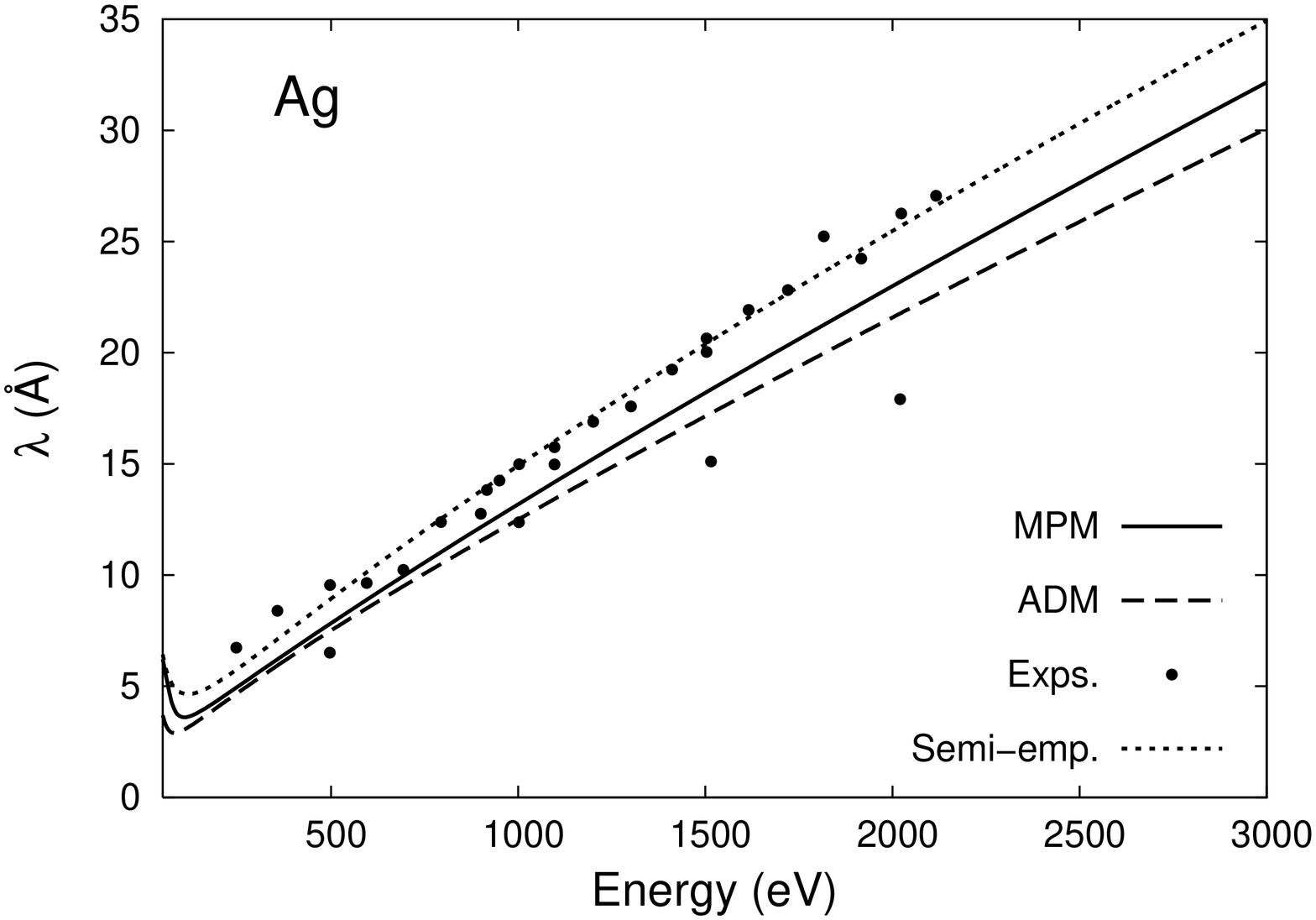}
\includegraphics[scale=0.35,angle=270]{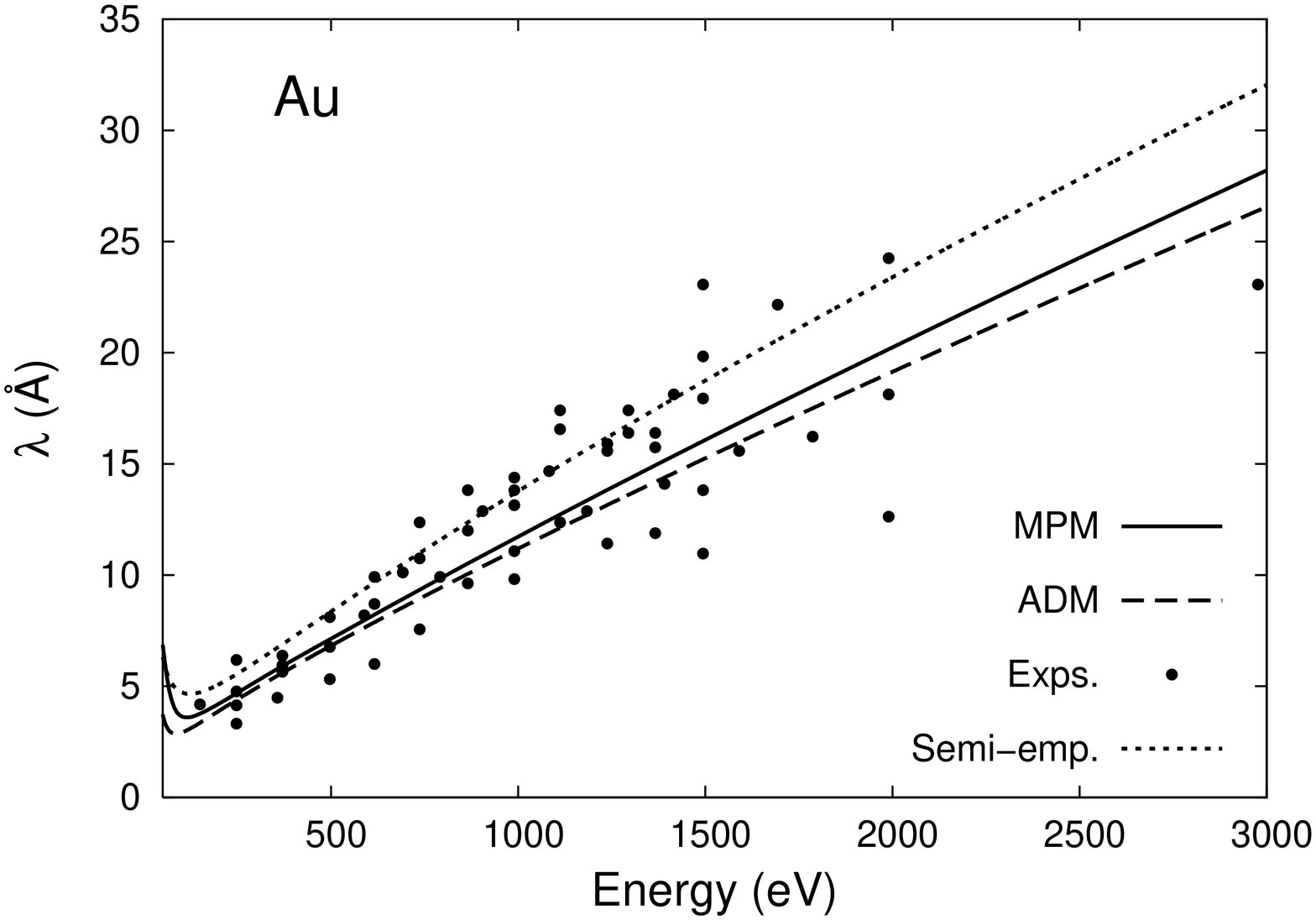}
\caption{\label{fig:imfp_ag}Inelastic mean free paths for a) silver (upper) and
b) gold (lower) calculated using the same {\it ab initio} dielectric function
as the basis for two different theoretical models: the many-pole self-energy
(MPM) model of Eq.~(\ref{IMFP2}), and the ``{\it ab initio} data" model (ADM)
described in the introduction. The theoretical results are compared to 
a semi-emperical curve \cite{powell99} and to multiple experimental data sets.
The references for the Exps.~are given in Ref.~\cite{powell99}. 
The theoretical models are plotted over the expected
range of validity of the semi-emperical curve.}
\end{figure}

\section{IV. Inelastic Mean Free Path}

We first calculate Eq.~(\ref{IMFP1})
for the IMFP in terms of the excited state
{\it self-energy} $\Sigma(E)$ of the fast electron.
\begin{equation}\label{IMFP2}
\lambda(E)=\sqrt{\frac{E}{2}}\frac{1}{\left|\im{\Sigma(E)}\right|}\;.
\end{equation}
Eq.~(\ref{IMFP2}) for the IMFP is consistent with the
decay of a single electron
wavefunction whose time dependence is given by $e^{-iE(p)t}$ \cite{quinn62}.
Eq.~(\ref{IMFP2}) is also equivalent to Eq.~(\ref{IMFP1}), because
the self-energy is proportional to the forward scattering amplitude, i.e.,
\begin{equation}\label{imimfp}
\im{\Sigma({\bf  p})}=- {2\pi} n_a\, \im{f({\bf p,p})}\;,
\end{equation}
i.e., the equivalence of Eq.~(\ref{IMFP1}) and Eq.~(\ref{IMFP2}) follows
from the optical theorem.

%

The explicit dependence of the self-energy on the 
dielectric function in Eq.~(\ref{scrncoul}), the many-pole model
of Eq.~(\ref{epsJ}), and (\ref{lund}), and
the full-spectrum FEFF8OP code are all that are needed to carry out
{\it ab initio} calculations of IMFPs according to Eq.~(\ref{IMFP2}).
Our many-pole model IMFP calculations (labeled MPM) are 
shown for several materials in
Figs.~\ref{fig:imfp_cu} and \ref{fig:imfp_ag},
together with best fits \cite{powell99} to currently available data. 
The fit lines in Figs.~\ref{fig:imfp_cu} and \ref{fig:imfp_ag} are
based on multiple data sets which were taken up to $3000$ eV and
are expected to accurately describe the IMFP as low as $50$ eV. 
Fig.~\ref{fig:imfp_ag}
also shows a calculation (labeled ``ADM") which uses our 
{\it ab initio} $\epsilon(\omega)$ as input data to the
semi-empirical optical-data model  of
Ref.~\cite{salvat05}. Note that the MPM and ADM results are in
in good agreement with each other, which verifies that the different 
extensions to finite $q$ discussed in Sec.\ II.\ do not lead to 
significantly different results. 
Both theoretical models are plotted here over
the expected range of validity of the fit line, but can be extended
with our codes to energies up to about 100 keV. Although the
agreement with experiment is reasonable, our calculations tend
to underestimate the experimental IMFP somewhat for high $Z$
materials.
%
%
%


\section{V. Stopping Power}

\begin{figure}
\includegraphics[scale=0.35,angle=270]{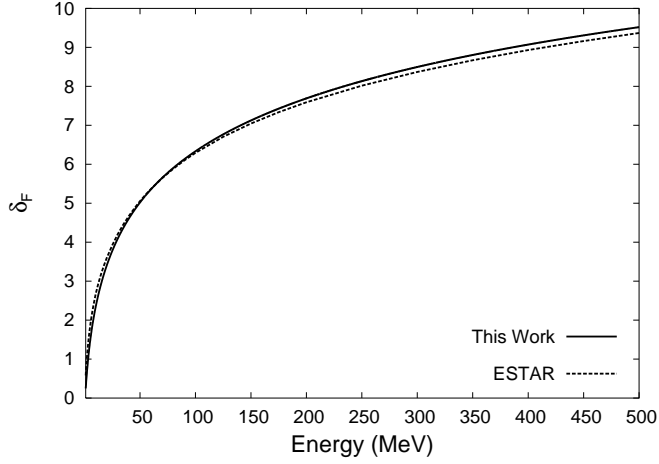}
\caption{\label{fig:deltas}Fano's density effect correction to the stopping 
power from Eq.~(\ref{sp}) as calculated in this work (solid), and compared to
semi-empirical values \cite{estar05} for copper (dashes).}
\end{figure}

%
As noted in the introduction the CSP is
the net reaction force $S(E) = -dE/dx$ due to electronic collisions
at a given energy $E$ that slows a fast probe electron. Over the 
range of energies from about 10 eV up to about 10 MeV
the CSP is the main
contribution to the total stopping power.  Above this
energy the total stopping power may be dominated by bremsstrahlung.
The CSP is calculated in this work using Eq.~(\ref{SP1}), where the DCS is
related to our {\it ab initio} loss function using the formulation of 
Ref.~\cite{salvat05}.  This model is thought to be applicable with confidence
for energies above about $100$ eV, and appears to be applicable as low
as about $10$ eV. However, it is not obvious why a model
based on the first Born approximation should be valid 
at such low energies. In the relativistic limit Eq.~(\ref{SP1}) reduces to the 
well-known Bethe-formula \cite{bethe30,inokuti71,fano63,fermi40}
for the stopping power
\begin{equation}\label{sp}
S(E)=\frac{\pi}{2}\frac{\Omega_p^2}{v^2}\left[
\ln{\left(\frac{E^2}{I^2}\frac{\gamma+1}{2}\right)}
+F(\gamma)
-\delta_F(\gamma) \right]\, ,
\end{equation}
%
where $\gamma=(1-v^2/c^2)^{-1/2}$ is the relativistic dilation factor, and 
$F(\gamma)$ is given by
\begin{equation} 
F(\gamma)=\left[\frac{1}{\gamma^2}-\frac{2\gamma-1}{\gamma^2}\ln{2}
+\frac{1}{8}{\left(\frac{\gamma-1}{\gamma}\right)}^2\right]\;.
\end{equation}
Also appearing in Eq.~(\ref{sp}) are
the ``mean excitation energy" $I$ defined in Sec.~II.,
and Fano's density correction \cite{fano63} $\delta_F$.
The density correction $\delta_F$ is due solely to transverse interactions,
and can be neglected for non-relativistic particles. A detailed
description of how $\delta_F$ can be calculated as a functional of
the loss function, is given in Ref.~\cite{salvat05}.
%
%
%
%
%
Fig.~\ref{fig:deltas} shows the density correction $\delta_F(E)$ 
for copper as calculated using 
our {\it ab initio} dielectric function, and for comparison, 
the semi-empirical values used by ESTAR \cite{estar05}. 
ESTAR is an on-line 
implementation of Eq.~(\ref{sp}) which 
semi-empirical values of $I$ as input. The mean excitation energy and the 
density correction have, heretofore, been 
difficult to calculate from first-principles, as they require accurate 
values of the OOS over a very large energy spectrum. However, our 
full-spectrum approach clearly gives reasonable agreement with experiment.

\begin{figure}
\includegraphics[scale=0.35,angle=270]{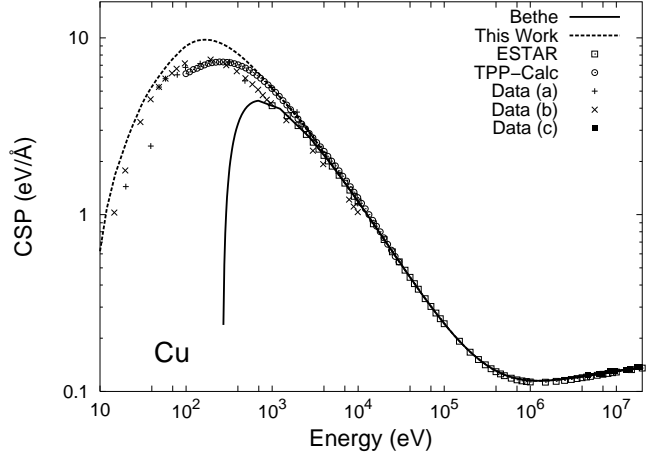}
\caption{
\label{fig:sp_cu}Collision stopping power for copper as
calculated using the {\it ab initio} dielectric function of this work (dashes)
in the ADM (see text).  Also shown are semi-empirical values of the CSP
from ESTAR\cite{estar05} (open squares), and semi-empirical CSP values
(labeled TPP-calc) based on the Penn model \cite{powell05} (circles),
and CSP values from experimental data: (+) \cite{luo91}, (x) \cite{hov96},
and (solid squares) \cite{mac98}.}
\end{figure}

For relativistic probe particles, the excellent agreement
of the Bethe formula in Eq.~(\ref{sp}) for the CSP 
is well known, so we have included data from ESTAR in lieu of experiment
where necessary. The difference between Eq.~(\ref{sp}) and Eq.~(\ref{SP1})
only appears in the non-relativistic regime, and can be seen in 
Fig.~\ref{fig:sp_cu}, where Eq.~(\ref{sp}) begins to fail around $5000$ eV.
%
%
\begin{figure}
\includegraphics[scale=0.35,angle=270]{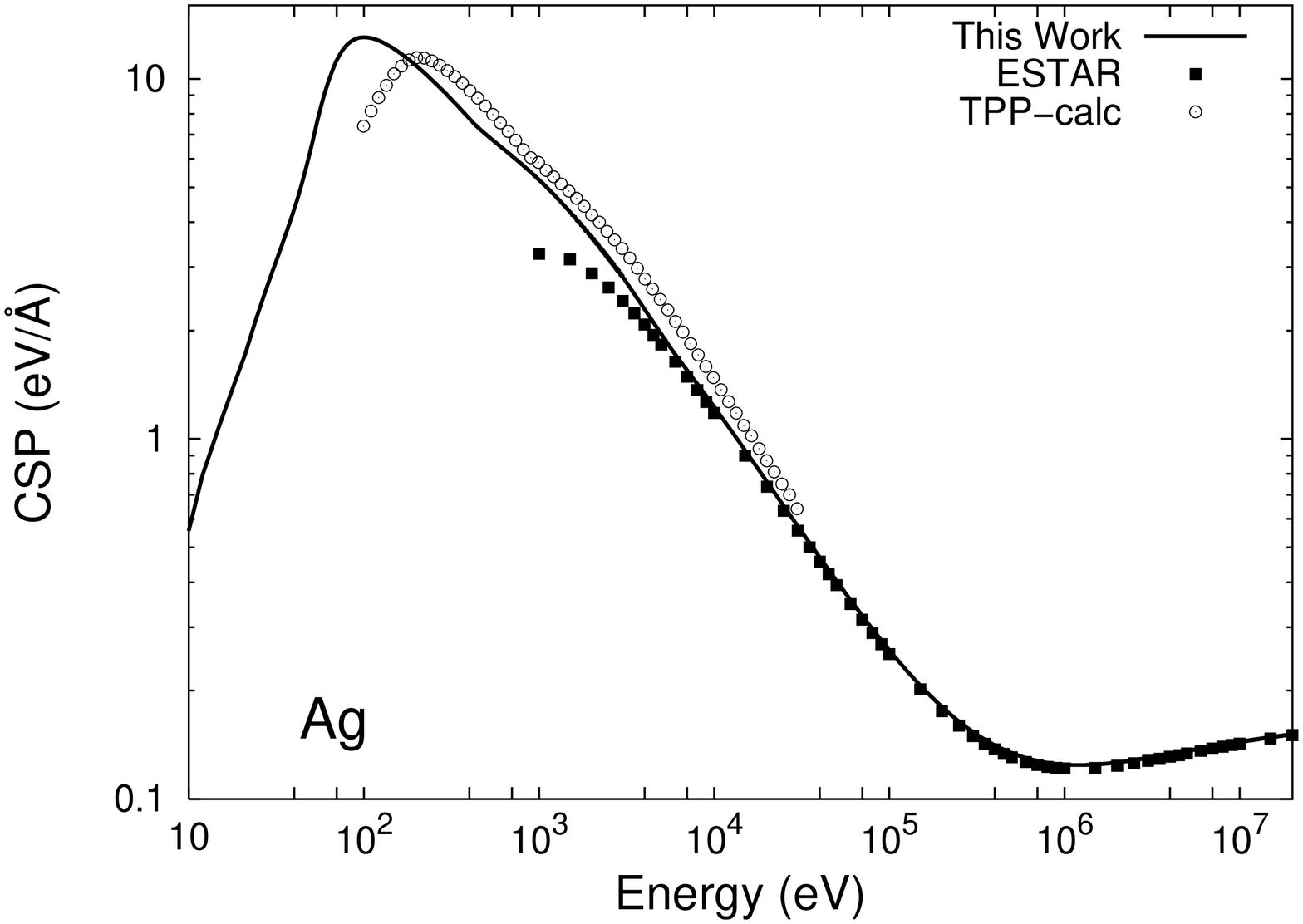}
\includegraphics[scale=0.35,angle=270]{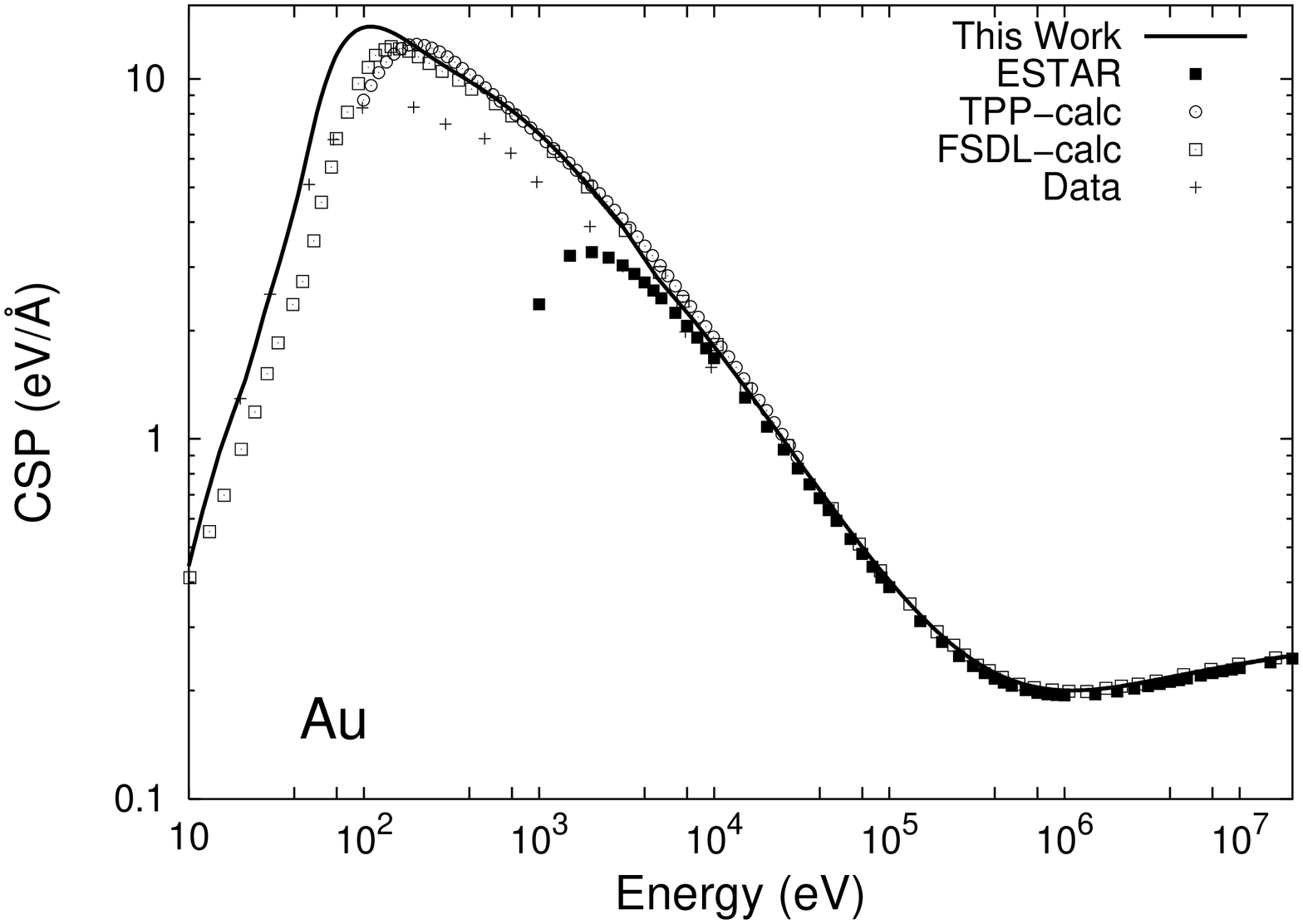}
\caption{\label{fig:sp_Au}Collision stopping powers for a) Ag (upper) and b) Au
(lower), with labels as in Fig.~\ref{fig:sp_cu}. Also shown for Au are the
semi-empirical CSP values as calculated in Ref.~\cite{salvat05} (labelled FSDL-calc), and CSP values
from experiment \cite{luo91}.}
\end{figure}
%
In order to calculate CSPs that are in good agreement with experiment 
for both non-relativistic and relativistic probe electrons
we apply Eq.~(\ref{SP1}) with the more general form of $d\sigma/d\omega$
given in Ref.~{\cite{salvat05}}, but using our calculated
dielectric function as input.
Figs.~\ref{fig:sp_cu} and \ref{fig:sp_Au}
show that the calculated CSPs for high energy electrons
for the systems studied here (Cu, Ag and Au)
are in good agreement with the ESTAR results. For lower energies the
CSP calculations of this work show significantly better agreement with
data than the Bethe formula.

\begin{figure}
\includegraphics[scale=0.35,angle=270]{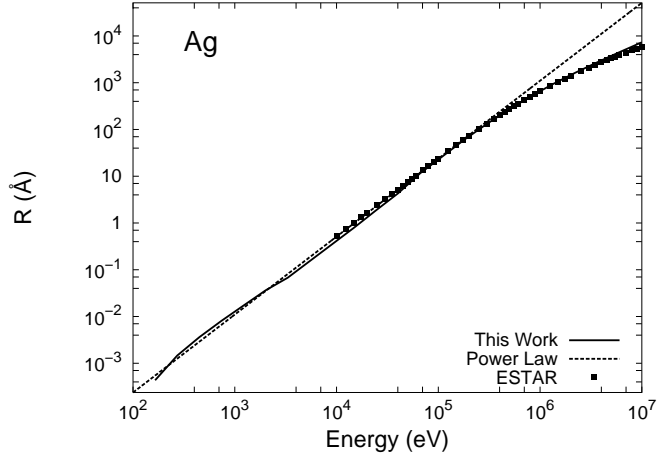}
\caption{\label{fig:range}
The range $R(E)$ as given by Eq.~(\ref{eq:range}) of electrons in silver as 
calculated in this work (solid) and compared to ESTAR (squares) and to
a pure power-law $R(E)=0.271 {E}^{5/3}$ (dashes) with $E$ in Hartrees.
}
\end{figure}

It is interesting to note that the CSP falls off approximately as
a power-law of the energy over a few decades beyond the peak loss,
but before the transverse effects take over. 
Because of this, the range $R(E)$ defined by
\begin{equation}\label{eq:range}
R(E)=\int_{0}^{E}\frac{dE}{S(E)},
\end{equation}
is also well approximated as a power-law. This approximate
power-law dependence of $R(E)$ is shown in Fig.~\ref{fig:range} for
silver, and compared with our full calculation and ESTAR.

\section{VI. Conclusions}
We have presented a general real-space Green's function approach
for {\it ab initio} calculations of
inelastic losses and stopping powers in condensed matter. Unlike
most current approaches, our method is based on {\it ab initio}
calculations of dielectric response, and does not rely on empirical
optical data.
We find that our many-pole self-energy model, which is derived from 
our many-pole dielectric function, yields
inelastic mean free paths in better agreement with experimental data than 
the single-pole plasmon model. Thus our many-pole self-energy 
more accurately accounts for inelastic losses in various electron and x-ray
spectroscopies than the single plasmon-pole self-energy.
We also find that accurate calculations of inelastic losses depend
primarily on the
quality of the calculated $q=0$ dielectric function, and not on the
details of the extension to finite $q$. 
Using our {\it ab initio} dielectric function, we also calculate the
mean excitation energy and thus  stopping powers for relativistic electrons,
obtaining results in good agreement with experimental data. Furthermore,
using the {\it ab initio} ADM we can extend the stopping power calculation
down to energies of $O(10)$ eV, i.e., much lower than the Bethe formula,
 while still maintaining reasonable agreement with experiment. 
Our approach for calculating stopping powers
of electrons can be extended to protons and other ions by suitably
modifying the kinematics. We have also found that the net range (or 
path length) $R(E)$
can be represented as a power law over several decades of energy.
That such a simple analytic description of the range of probe electrons
in a solid exists is interesting and also could be useful for reducing
computation times in Monte Carlo calculations used to study radiation
damage and in other applications where numerous electron energy-loss tracks
need to be considered. 
In conclusion, we believe our approach has the potential to complement or
provide an alternative to semi-empirical approaches for calculations of
IMFPs and stopping powers in condensed matter.

\section{Acknowledgments}
We wish to thank G. Bertsch, H. Bichsel, J. Fern\'andez-Varea, 
C. Powell, P. Rez, and E. Stern for comments and suggestions.
This work is supported in part by the DOE Grant DE-FG03-97ER45623
(JJR) NIH NCRR BTP Grant RR-01209 (JJK), and NIST Grant 70 NAMB 2H003
(APS) and was facilitated by the
DOE Computational Materials Science Network.

\begin{thebibliography}{36}
\expandafter\ifx\csname natexlab\endcsname\relax\def\natexlab#1{#1}\fi
\expandafter\ifx\csname bibnamefont\endcsname\relax
  \def\bibnamefont#1{#1}\fi
\expandafter\ifx\csname bibfnamefont\endcsname\relax
  \def\bibfnamefont#1{#1}\fi
\expandafter\ifx\csname citenamefont\endcsname\relax
  \def\citenamefont#1{#1}\fi
\expandafter\ifx\csname url\endcsname\relax
  \def\url#1{\texttt{#1}}\fi
\expandafter\ifx\csname urlprefix\endcsname\relax\def\urlprefix{URL }\fi
\providecommand{\bibinfo}[2]{#2}
\providecommand{\eprint}[2][]{\url{#2}}

\bibitem[{\citenamefont{Bethe}(1930)}]{bethe30}
\bibinfo{author}{\bibfnamefont{H.~A.} \bibnamefont{Bethe}},
  \bibinfo{journal}{Ann. Phys.} \textbf{\bibinfo{volume}{5}},
  \bibinfo{pages}{325} (\bibinfo{year}{1930}), \bibinfo{note}{this paper is
  reviewed in Ref. \cite{inokuti71}}.

\bibitem[{\citenamefont{Fermi}(1940)}]{fermi40}
\bibinfo{author}{\bibfnamefont{E.}~\bibnamefont{Fermi}},
  \bibinfo{journal}{Phys. Rev.} \textbf{\bibinfo{volume}{57}},
  \bibinfo{pages}{485} (\bibinfo{year}{1940}).

\bibitem[{\citenamefont{Fano}(1963)}]{fano63}
\bibinfo{author}{\bibfnamefont{U.}~\bibnamefont{Fano}}, \bibinfo{journal}{Ann.
  Rev. Nucl. Sci.} \textbf{\bibinfo{volume}{13}}, \bibinfo{pages}{1}
  (\bibinfo{year}{1963}).

\bibitem[{\citenamefont{Fern\'andez-Varea
  et~al.}(2005)\citenamefont{Fern\'andez-Varea, Salvat, Dingfelder, and
  Liljequist}}]{salvat05}
\bibinfo{author}{\bibfnamefont{J.~M.} \bibnamefont{Fern\'andez-Varea}},
  \bibinfo{author}{\bibfnamefont{F.}~\bibnamefont{Salvat}},
  \bibinfo{author}{\bibfnamefont{M.}~\bibnamefont{Dingfelder}},
  \bibnamefont{and}
  \bibinfo{author}{\bibfnamefont{D.}~\bibnamefont{Liljequist}},
  \bibinfo{journal}{Nucl. Instr. and Meth. B} \textbf{\bibinfo{volume}{229}},
  \bibinfo{pages}{187} (\bibinfo{year}{2005}).

\bibitem[{\citenamefont{Powell and Jablonski}(1999)}]{powell99}
\bibinfo{author}{\bibfnamefont{C.~J.} \bibnamefont{Powell}} \bibnamefont{and}
  \bibinfo{author}{\bibfnamefont{A.}~\bibnamefont{Jablonski}},
  \bibinfo{journal}{J. Phys. Chem. Ref. Data} \textbf{\bibinfo{volume}{28}},
  \bibinfo{pages}{19} (\bibinfo{year}{1999}).

\bibitem[{\citenamefont{Bichsel}(1996)}]{bichsel93}
\bibinfo{author}{\bibfnamefont{H.}~\bibnamefont{Bichsel}}, in
  \emph{\bibinfo{booktitle}{Atomic, Moecular, and Optical Physics Handbook}},
  edited by \bibinfo{editor}{\bibfnamefont{G.~W.~F.} \bibnamefont{Drake}}
  (\bibinfo{publisher}{AIP Press, Woodbury, N.Y.}, \bibinfo{year}{1996}).

\bibitem[{\citenamefont{Campillo et~al.}(1999)\citenamefont{Campillo, Pitarke,
  Rubio, Zarate, and Echenique}}]{rubio99}
\bibinfo{author}{\bibfnamefont{I.}~\bibnamefont{Campillo}},
  \bibinfo{author}{\bibfnamefont{J.~M.} \bibnamefont{Pitarke}},
  \bibinfo{author}{\bibfnamefont{A.}~\bibnamefont{Rubio}},
  \bibinfo{author}{\bibfnamefont{E.}~\bibnamefont{Zarate}}, \bibnamefont{and}
  \bibinfo{author}{\bibfnamefont{P.~M.} \bibnamefont{Echenique}},
  \bibinfo{journal}{Phys. Rev. Lett.} \textbf{\bibinfo{volume}{83}},
  \bibinfo{pages}{2230} (\bibinfo{year}{1999}).

\bibitem[{\citenamefont{Soininen et~al.}(2003)\citenamefont{Soininen, Rehr, and
  Shirley}}]{soininen03}
\bibinfo{author}{\bibfnamefont{J.~A.} \bibnamefont{Soininen}},
  \bibinfo{author}{\bibfnamefont{J.~J.} \bibnamefont{Rehr}}, \bibnamefont{and}
  \bibinfo{author}{\bibfnamefont{E.~L.} \bibnamefont{Shirley}},
  \bibinfo{journal}{J. Phys.: Condens. Matter} \textbf{\bibinfo{volume}{15}},
  \bibinfo{pages}{2572} (\bibinfo{year}{2003}).

\bibitem[{\citenamefont{Penn}(1987)}]{dpenn87}
\bibinfo{author}{\bibfnamefont{D.~R.} \bibnamefont{Penn}},
  \bibinfo{journal}{Phys. Rev. B} \textbf{\bibinfo{volume}{35}},
  \bibinfo{pages}{482} (\bibinfo{year}{1987}).

\bibitem[{\citenamefont{Sternheimer et~al.}(1982)\citenamefont{Sternheimer,
  Seltzer, and Berger}}]{sternheimer82}
\bibinfo{author}{\bibfnamefont{R.~M.} \bibnamefont{Sternheimer}},
  \bibinfo{author}{\bibfnamefont{S.~M.} \bibnamefont{Seltzer}},
  \bibnamefont{and} \bibinfo{author}{\bibfnamefont{M.~J.}
  \bibnamefont{Berger}}, \bibinfo{journal}{Phys. Rev. B}
  \textbf{\bibinfo{volume}{26}}, \bibinfo{pages}{6067} (\bibinfo{year}{1982}).

\bibitem[{\citenamefont{Rehr et~al.}()\citenamefont{Rehr, Kas, Prange, Vila,
  Ankudinov, Campbell, and Sorini}}]{rehr06}
\bibinfo{author}{\bibfnamefont{J.~J.} \bibnamefont{Rehr}},
  \bibinfo{author}{\bibfnamefont{J.~J.} \bibnamefont{Kas}},
  \bibinfo{author}{\bibfnamefont{M.~P.} \bibnamefont{Prange}},
  \bibinfo{author}{\bibfnamefont{F.~D.} \bibnamefont{Vila}},
  \bibinfo{author}{\bibfnamefont{A.~L.} \bibnamefont{Ankudinov}},
  \bibinfo{author}{\bibfnamefont{L.~W.} \bibnamefont{Campbell}},
  \bibnamefont{and} \bibinfo{author}{\bibfnamefont{A.~P.}
  \bibnamefont{Sorini}}, \bibinfo{note}{arXiv:cond-mat/0601242, 2006,
  unpublished.}

\bibitem[{\citenamefont{Ankudinov et~al.}(1998)\citenamefont{Ankudinov, Ravel,
  Rehr, and Conradson}}]{ankudinov98}
\bibinfo{author}{\bibfnamefont{A.~L.} \bibnamefont{Ankudinov}},
  \bibinfo{author}{\bibfnamefont{B.}~\bibnamefont{Ravel}},
  \bibinfo{author}{\bibfnamefont{J.~J.} \bibnamefont{Rehr}}, \bibnamefont{and}
  \bibinfo{author}{\bibfnamefont{S.~D.} \bibnamefont{Conradson}},
  \bibinfo{journal}{Phys. Rev. B} \textbf{\bibinfo{volume}{58}},
  \bibinfo{pages}{7565} (\bibinfo{year}{1998}).

\bibitem[{\citenamefont{Prange et~al.}()\citenamefont{Prange, Rivas, Rehr, and
  Ankudinov}}]{prangeetal05}
\bibinfo{author}{\bibfnamefont{M.}~\bibnamefont{Prange}},
  \bibinfo{author}{\bibfnamefont{G.}~\bibnamefont{Rivas}},
  \bibinfo{author}{\bibfnamefont{J.}~\bibnamefont{Rehr}}, \bibnamefont{and}
  \bibinfo{author}{\bibfnamefont{A.}~\bibnamefont{Ankudinov}},
  \bibinfo{note}{unpublished.}

\bibitem[{\citenamefont{Liljequist}(1983)}]{lilj83}
\bibinfo{author}{\bibfnamefont{D.}~\bibnamefont{Liljequist}},
  \bibinfo{journal}{J. Phys. D: Appl. Phys.} \textbf{\bibinfo{volume}{16}},
  \bibinfo{pages}{1567} (\bibinfo{year}{1983}).

\bibitem[{\citenamefont{Ashley et~al.}(1979)\citenamefont{Ashley, Cowan,
  Ritchie, Anderson, and Hoelzl}}]{ashley79}
\bibinfo{author}{\bibfnamefont{J.~C.} \bibnamefont{Ashley}},
  \bibinfo{author}{\bibfnamefont{J.~J.} \bibnamefont{Cowan}},
  \bibinfo{author}{\bibfnamefont{R.~H.} \bibnamefont{Ritchie}},
  \bibinfo{author}{\bibfnamefont{V.~E.} \bibnamefont{Anderson}},
  \bibnamefont{and} \bibinfo{author}{\bibfnamefont{J.}~\bibnamefont{Hoelzl}},
  \bibinfo{journal}{Thin Solid Films} \textbf{\bibinfo{volume}{60}},
  \bibinfo{pages}{361} (\bibinfo{year}{1979}).

\bibitem[{\citenamefont{Ashley}(1988)}]{ashley88}
\bibinfo{author}{\bibfnamefont{J.~C.} \bibnamefont{Ashley}},
  \bibinfo{journal}{J. Electron Spectrosc. Relat. Phenom.}
  \textbf{\bibinfo{volume}{46}}, \bibinfo{pages}{199} (\bibinfo{year}{1988}).

\bibitem[{\citenamefont{Prange et~al.}(2005)\citenamefont{Prange, Rivas, and
  Rehr}}]{prangetables05}
\bibinfo{author}{\bibfnamefont{M.}~\bibnamefont{Prange}},
  \bibinfo{author}{\bibfnamefont{G.}~\bibnamefont{Rivas}}, \bibnamefont{and}
  \bibinfo{author}{\bibfnamefont{J.~J.} \bibnamefont{Rehr}},
  \emph{\bibinfo{title}{Table of Optical Constants for Mg, Al, Cu, Ag, Au, Bi
  and C.}} (\bibinfo{publisher}{World Wide Web},
  \bibinfo{address}{\url{http://leonardo.phys.washington.edu/feff/opcons/}},
  \bibinfo{year}{2005}).

\bibitem[{\citenamefont{Hagemann et~al.}(1975)\citenamefont{Hagemann, Gudat,
  and Kunz}}]{hagemann75}
\bibinfo{author}{\bibfnamefont{H.~J.} \bibnamefont{Hagemann}},
  \bibinfo{author}{\bibfnamefont{W.}~\bibnamefont{Gudat}}, \bibnamefont{and}
  \bibinfo{author}{\bibfnamefont{C.}~\bibnamefont{Kunz}}, \bibinfo{journal}{J.
  Opt. Soc. Am.} \textbf{\bibinfo{volume}{65}}, \bibinfo{pages}{7421}
  (\bibinfo{year}{1975}).

\bibitem[{\citenamefont{Hagemann et~al.}(1974)\citenamefont{Hagemann, Gudat,
  and Kunz}}]{hagemann74}
\bibinfo{author}{\bibfnamefont{H.~J.} \bibnamefont{Hagemann}},
  \bibinfo{author}{\bibfnamefont{W.}~\bibnamefont{Gudat}}, \bibnamefont{and}
  \bibinfo{author}{\bibfnamefont{C.}~\bibnamefont{Kunz}},
  \emph{\bibinfo{title}{Optical Constants from the Far Infrared to the X-ray
  Region: Mg, Al, Cu, Ag, Au, Bi, C and Al$_2$O$_3$, DESY SR-7417}}
  (\bibinfo{publisher}{Desy, Hamburg, W. Germany}, \bibinfo{year}{1974}).

\bibitem[{\citenamefont{Sorini et~al.}(2005)\citenamefont{Sorini, Kas, Rehr,
  and Prange}}]{sorinitables06}
\bibinfo{author}{\bibfnamefont{A.~P.} \bibnamefont{Sorini}},
  \bibinfo{author}{\bibfnamefont{J.}~\bibnamefont{Kas}},
  \bibinfo{author}{\bibfnamefont{J.~J.} \bibnamefont{Rehr}}, \bibnamefont{and}
  \bibinfo{author}{\bibfnamefont{M.~P.} \bibnamefont{Prange}},
  \emph{\bibinfo{title}{Tables of mean free paths and stopping powers.}}
  (\bibinfo{publisher}{World Wide Web},
  \bibinfo{address}{\url{http://leonardo.phys.washington.edu/feff/loss/}},
  \bibinfo{year}{2005}).

\bibitem[{\citenamefont{Tschalar and Bichsel}(1968)}]{bichsel68}
\bibinfo{author}{\bibfnamefont{C.}~\bibnamefont{Tschalar}} \bibnamefont{and}
  \bibinfo{author}{\bibfnamefont{H.}~\bibnamefont{Bichsel}},
  \bibinfo{journal}{Phys. Rev.} \textbf{\bibinfo{volume}{175}},
  \bibinfo{pages}{476} (\bibinfo{year}{1968}).

\bibitem[{\citenamefont{ICRU}(1984)}]{icru84}
\bibinfo{author}{\bibnamefont{ICRU}}, \emph{\bibinfo{title}{ICRU Report 37,
  Stopping Powers and Ranges for Protons and Alpha Particles.}}
  (\bibinfo{publisher}{International Commission of Radiation Units and
  Measurements.}, \bibinfo{year}{1984}).

\bibitem[{\citenamefont{Bichsel}(1992)}]{bichsel92}
\bibinfo{author}{\bibfnamefont{H.}~\bibnamefont{Bichsel}},
  \bibinfo{journal}{Phys. Rev. A} \textbf{\bibinfo{volume}{46}},
  \bibinfo{pages}{5761} (\bibinfo{year}{1992}).

\bibitem[{\citenamefont{Lundqvist}(1967)}]{lundqvist67}
\bibinfo{author}{\bibfnamefont{B.}~\bibnamefont{Lundqvist}},
  \bibinfo{journal}{Phys. Kondens. Materie} \textbf{\bibinfo{volume}{6}},
  \bibinfo{pages}{193} (\bibinfo{year}{1967}).

\bibitem[{\citenamefont{Kas et~al.}()\citenamefont{Kas, Sorini, Prange, and
  Rehr}}]{kasetal05}
\bibinfo{author}{\bibfnamefont{J.}~\bibnamefont{Kas}},
  \bibinfo{author}{\bibfnamefont{A.}~\bibnamefont{Sorini}},
  \bibinfo{author}{\bibfnamefont{M.}~\bibnamefont{Prange}}, \bibnamefont{and}
  \bibinfo{author}{\bibfnamefont{J.}~\bibnamefont{Rehr}},
  \bibinfo{note}{unpublished.}

\bibitem[{\citenamefont{Salvat et~al.}(2003)\citenamefont{Salvat,
  Fern\'andez-Varea, and Sempau}}]{penelope}
\bibinfo{author}{\bibfnamefont{F.}~\bibnamefont{Salvat}},
  \bibinfo{author}{\bibfnamefont{J.~M.} \bibnamefont{Fern\'andez-Varea}},
  \bibnamefont{and} \bibinfo{author}{\bibfnamefont{J.}~\bibnamefont{Sempau}},
  \emph{\bibinfo{title}{PENELOPE - A Code System for Monte Carlo Simulation of
  Electron and Photon Transport}} (\bibinfo{publisher}{OECD/Nuclear Energy
  Agency, Issy-les-Moulineaux, France},
  \bibinfo{address}{\url{http://www.nea.fr}}, \bibinfo{year}{2003}).

\bibitem[{\citenamefont{Soininen et~al.}(2005)\citenamefont{Soininen,
  Ankudinov, and Rehr}}]{soininen05}
\bibinfo{author}{\bibfnamefont{J.~A.} \bibnamefont{Soininen}},
  \bibinfo{author}{\bibfnamefont{A.~L.} \bibnamefont{Ankudinov}},
  \bibnamefont{and} \bibinfo{author}{\bibfnamefont{J.~J.} \bibnamefont{Rehr}},
  \bibinfo{journal}{Phys. Rev. B} \textbf{\bibinfo{volume}{72}},
  \bibinfo{pages}{045136} (\bibinfo{year}{2005}).

\bibitem[{\citenamefont{Hedin and Lundqvist}(1969)}]{hedin69}
\bibinfo{author}{\bibfnamefont{L.}~\bibnamefont{Hedin}} \bibnamefont{and}
  \bibinfo{author}{\bibfnamefont{S.}~\bibnamefont{Lundqvist}},
  \bibinfo{journal}{Solid State Phys.} \textbf{\bibinfo{volume}{23}},
  \bibinfo{pages}{1} (\bibinfo{year}{1969}).

\bibitem[{\citenamefont{Werner}(2001)}]{werner}
\bibinfo{author}{\bibfnamefont{W.~S.~M.} \bibnamefont{Werner}},
  \bibinfo{journal}{Surf. Interface Anal.} \textbf{\bibinfo{volume}{31}},
  \bibinfo{pages}{141} (\bibinfo{year}{2001}).

\bibitem[{\citenamefont{Tanuma et~al.}(2005)\citenamefont{Tanuma, Powell, and
  Penn}}]{powell05}
\bibinfo{author}{\bibfnamefont{S.}~\bibnamefont{Tanuma}},
  \bibinfo{author}{\bibfnamefont{C.~J.} \bibnamefont{Powell}},
  \bibnamefont{and} \bibinfo{author}{\bibfnamefont{D.~R.} \bibnamefont{Penn}},
  \bibinfo{journal}{Surf. Interface Anal.} \textbf{\bibinfo{volume}{37}},
  \bibinfo{pages}{978} (\bibinfo{year}{2005}).

\bibitem[{\citenamefont{Quinn}(1962)}]{quinn62}
\bibinfo{author}{\bibfnamefont{J.~J.} \bibnamefont{Quinn}},
  \bibinfo{journal}{Phys. Rev.} \textbf{\bibinfo{volume}{126}},
  \bibinfo{pages}{1453} (\bibinfo{year}{1962}).

\bibitem[{\citenamefont{Inokuti}(1971)}]{inokuti71}
\bibinfo{author}{\bibfnamefont{M.}~\bibnamefont{Inokuti}},
  \bibinfo{journal}{Rev. Mod. Phys.} \textbf{\bibinfo{volume}{43}},
  \bibinfo{pages}{297} (\bibinfo{year}{1971}).

\bibitem[{\citenamefont{Berger et~al.}(2005)\citenamefont{Berger, Coursey,
  Zucker, and Chang}}]{estar05}
\bibinfo{author}{\bibfnamefont{M.~J.} \bibnamefont{Berger}},
  \bibinfo{author}{\bibfnamefont{J.~S.} \bibnamefont{Coursey}},
  \bibinfo{author}{\bibfnamefont{M.~A.} \bibnamefont{Zucker}},
  \bibnamefont{and} \bibinfo{author}{\bibfnamefont{J.}~\bibnamefont{Chang}},
  \emph{\bibinfo{title}{ESTAR, PSTAR, and ASTAR: Computer Programs for
  Calculating Stopping-Power and Range Tables for Electrons, Protons, and
  Helium Ions (version 1.2.3)}} (\bibinfo{publisher}{National Institute of
  Standards and Technology, Gaithersburd, MD.},
  \bibinfo{address}{\url{http://physics.nist.gov/Star}}, \bibinfo{year}{2005}).

\bibitem[{\citenamefont{Luo et~al.}(1991)\citenamefont{Luo, Zhang, and
  Joy}}]{luo91}
\bibinfo{author}{\bibfnamefont{S.}~\bibnamefont{Luo}},
  \bibinfo{author}{\bibfnamefont{X.}~\bibnamefont{Zhang}}, \bibnamefont{and}
  \bibinfo{author}{\bibfnamefont{D.~C.} \bibnamefont{Joy}},
  \bibinfo{journal}{Radiat. Eff. Def. Solids} \textbf{\bibinfo{volume}{117}},
  \bibinfo{pages}{235} (\bibinfo{year}{1991}).

\bibitem[{\citenamefont{Hovington et~al.}(1996)\citenamefont{Hovington, Joy,
  Gauvin, and Evans}}]{hov96}
\bibinfo{author}{\bibfnamefont{P.}~\bibnamefont{Hovington}},
  \bibinfo{author}{\bibfnamefont{D.~C.} \bibnamefont{Joy}},
  \bibinfo{author}{\bibfnamefont{R.}~\bibnamefont{Gauvin}}, \bibnamefont{and}
  \bibinfo{author}{\bibfnamefont{N.}~\bibnamefont{Evans}},
  \bibinfo{journal}{Scanning Microsc.} \textbf{\bibinfo{volume}{10}},
  \bibinfo{pages}{653} (\bibinfo{year}{1996}).

\bibitem[{\citenamefont{Macpherson}(1998)}]{mac98}
\bibinfo{author}{\bibfnamefont{M.~S.} \bibnamefont{Macpherson}}, Ph.D. thesis,
  \bibinfo{school}{NRC Report PIRS-0626} (\bibinfo{year}{1998}).

\end{thebibliography}
\end{document}